\documentclass[prl,aps,twocolumn,showpacs]{revtex4}
\usepackage{epsfig}
\usepackage{amsmath}
\usepackage{amssymb}
\begin{document}
\title{Three body problem in a dilute Bose-Einstein condensate}
\author{Thorsten K\"ohler}
\affiliation{Clarendon Laboratory, Department of Physics, 
University of Oxford, Oxford, OX1 3PU, United Kingdom}
\date{\today}
\begin{abstract}
\noindent
We derive the explicit three body contact potential for a dilute condensed Bose
gas from microscopic theory. The three body coupling constant exhibits the 
general form predicted by T.~T.~Wu [Phys.~Rev.~{\bf 113}, 1390 (1959)] and is 
determined in terms of the amplitudes of two and three body collisions in 
vacuum. In the present form the coupling constant becomes accessible to 
quantitative studies which should provide the crucial link between 
few body collisions and the stability of condensates with attractive two 
body forces.  
\end{abstract}
\maketitle
\noindent
Despite their diluteness atomic Bose Einstein condensates (BECs) are 
significantly influenced by inter-atomic forces. The binary potential
enters their properties mainly through two body low energy scattering 
observables described by the $s$ wave scattering length $a$. Recent experiments
\cite{InouyeCornish} 
have provided the opportunity to tune the scattering length from the repulsive
($a>0$) through the ideal gas ($a\sim 0$) to the attractive ($a<0$) regime
where the gas finally becomes unstable with respect to collapse
\cite{Roberts,Donley}. 
With this new technique one can thus access a wide range of dilute gas 
parameters and study phenomena related to few body collisions in the gas.

Virtually all properties of BECs at zero temperature are described by a
nonlinear Schr\"odinger equation (NLSE) 
\cite{Dalfovo}
\begin{align}
i\hbar\frac{\partial}{\partial t}\Psi=H_{\rm 1B}\Psi+g_2|\Psi|^2\Psi
+g_3|\Psi|^4\Psi+\ldots
\label{NLSE}
\end{align}
which, in this extended form, depends on the Hamiltonian of a single trapped
atom as well as the two and three body coupling constants $g_2$ and $g_3$, 
respectively.
Three body scattering obviously becomes significant at large scattering 
lengths, through the increase of the dilute gas parameter 
$\eta=\sqrt{|\Psi|^2 |a|^3}$, but also in the ideal gas regime. 
Small scattering lengths of only a few {\AA}, e.g.,
are needed to produce stable condensates with attractive interactions 
and several thousands of atoms in present day atom traps. Two body scattering 
alone, however, seems not sufficient to explain the experimental criterion 
for stability against collapse of $^{85}$Rb
\cite{Roberts}
and a fit of Eq.~(\ref{NLSE}) to the data gives an attractive three body 
interaction with very reasonable coupling constants 
$|{\rm Re}g_3|/\hbar$ of the order of $10^{-27}$ cm$^6$/s
\cite{Gammal}.
In this way the corresponding three body mean field shift of the energy 
density, ${\rm Re}g_3|\Psi|^6/3$, should become directly accessible 
to experiments.

The determination of the three body coupling constant in a dilute BEC, 
and its associated mean field energy, have a long history of 
theoretical research in many body physics
\cite{Fetter}.
Already in 1959 Wu
\cite{Wu}
discovered the general form 
$g_3=16\pi\hbar^2a^4(4\pi-3\sqrt{3})\ln(C\eta^2)/m$ for a  
Bose gas of hard spheres. The constant $C$ in the argument of the
logarithm, however, remained undetermined and only recently was 
significant progress made. In 1999 Braaten and Nieto 
\cite{BraatenNieto}
used effective field theory to determine Wu's general result by 
taking into account low energy three body scattering. They estimated
the argument of the logarithm on the basis of general assumptions 
on the length scales set by the inter-atomic potential.
Braaten et al.~\cite{BraatenHammerHermans}
also studied the role of parameters of the potential apart 
from the scattering length,
through fits to many body Monte Carlo simulations
\cite{Giorgini},
as well as the properties of the effective three body coupling constant 
when the scattering length is large in comparison to the range of the potential
\cite{BraatenBulgac}.

To the best of our knowledge,
the three body coupling constant has not been derived from a 
microscopic description of the gas and expressed in terms of the
binary potential. In this letter we provide a microscopic theory of three body 
collisions in a BEC and determine $g_2$ and $g_3$ to second order in the
dilute gas parameter $\eta$.
The derivation is based on the exact quantum evolution of correlation 
functions, i.e.~quantum expectation values of products of single mode 
annihilation and creation operators of the form $\langle a_{\bf p}\rangle$, 
$\langle a_{{\bf p}_2}a_{{\bf p}_1}\rangle$, 
$\langle a_{\bf q}^\dagger a_{\bf p}\rangle$ etc.. References 
\cite{Fricke,TKKB} provide a general approach to transform the exact
infinite hierarchy of dynamic equations for correlation functions
into a more favorable form that allows for a systematic 
truncation in accordance with Wick's theorem.
The resulting set of dynamic equations for non commutative cumulants 
\cite{TKKB} lends itself to iteration. We shall apply this cumulant 
approach here to express the exact dynamic equation for the condensate wave 
function of a uniform gas through nonlinear terms in $\Psi$ and $\Psi^*$ and 
determine $g_2$ and $g_3$ in equilibrium. 

According to Ref.~\cite{TKKB}
the exact quantum evolution of the condensate wave function 
$\Psi_{\bf p}=\langle a_{\bf p}\rangle$ depends on the pair function
$\Phi_{{\bf p}_1 {\bf p}_2}=\langle a_{{\bf p}_2}a_{{\bf p}_1}\rangle-
\langle a_{{\bf p}_2}\rangle\langle a_{{\bf p}_1}\rangle$ and the third order
cumulant $\Lambda$ which is associated with the correlation function
$\langle a_{\bf q}^\dagger a_{{\bf p}_2} a_{{\bf p}_1}\rangle$. 
We shall first consider the uniform gas limit.
The dynamic equation for $\Psi$ then reads
\cite{transinvariance}
\begin{align}
\nonumber
&i\hbar\frac{\partial}{\partial t}\Psi(t)=
\frac{1}{\sqrt{2\pi\hbar}^3}\int d{\bf q}d{\bf p}
V({\bf p}_1)\Lambda({\bf q},{\bf p},t)\\
&+\int d{\bf p}
V({\bf p})
\left[
\Phi({\bf p},t)+
\sqrt{2\pi\hbar}^3\delta({\bf p})\Psi^2(t)
\right]\Psi^*(t),
\label{NLSEexact}
\end{align}
where the integration variables are chosen as center of mass and 
relative atomic momenta ${\bf q}={\bf p}_1+{\bf p}_2$ and 
${\bf p}=({\bf p}_2-{\bf p}_1)/2$, respectively, and
$V({\bf p})=\int d{\bf r}V({\bf r})
\exp(-i{\bf p}\cdot{\bf r}/\hbar)/(\sqrt{2\pi\hbar})^3$ is the Fourier
transform of the potential into momentum space. Additional terms 
\cite{TKKB}
that are proportional to the density matrix of the non condensed fraction,
$\Gamma_{{\bf p} {\bf q}}=\langle a_{\bf q}^\dagger a_{\bf p}\rangle-
\langle a_{\bf q}^\dagger\rangle\langle a_{\bf p}\rangle$, 
neither contribute to $g_2$ nor to $g_3$ and have been omitted 
in Eq.~(\ref{NLSEexact}).

The explicit dependence of Eq.~(\ref{NLSEexact}) 
on higher order cumulants can be eliminated 
successively by solving their dynamic equations formally and 
inserting the solutions into Eq.~(\ref{NLSEexact})
\cite{TKKB}. 
With this iterative procedure Eq.~(\ref{NLSEexact}) 
can be expressed exactly through nonlinear terms in $\Psi$, $\Psi^*$ 
and $\Gamma$ up to a given order and a remainder which contains only
higher order cumulants. To obtain the coupling 
constants $g_2$ and $g_3$ we shall determine all the nonlinear terms in 
$\Psi$ and $\Psi^*$ up to the fifth order.

\begin{figure}
\begin{center}
\epsfig{file=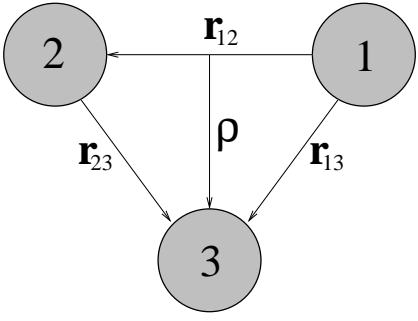,width=5.5cm}
\end{center}
\caption{Jacobi coordinates of three atoms. The spatial coordinates
are chosen as the vector from atom 1 to atom 2 (${\bf r}_{12}$) 
and the vector from the center of mass of atoms 1 and 2 to atom 3
($\boldsymbol{\rho}$).
The conjugate Jacobi momenta are ${\bf p}=({\bf p}_2-{\bf p}_1)/2$
and ${\bf q}=(2{\bf p}_3-{\bf p}_1-{\bf p}_2)/3$, respectively, where
${\bf p}_1,{\bf p}_2,{\bf p}_3$ denote the momenta of the single atoms.}
\label{fig:Jacobi}
\end{figure}

The formal solutions of the dynamic equations of $\Phi$ and $\Lambda$
\cite{TKKB}
involve the triple function $\chi$, i.e.~the third order cumulant associated
with the correlation function 
$\langle a_{{\bf p}_3} a_{{\bf p}_2} a_{{\bf p}_1}\rangle$, and 
the Green's function corresponding to two atoms described by the 
two body Hamiltonian 
$H_{\rm 2B}=-\frac{\hbar^2}{m}\Delta_{\bf r}+V({\bf r})$ 
in their center of mass frame. The relevant retarded Green's function 
is given by \cite{Newton} 
$G(t)=\frac{1}{i\hbar}\theta(t)\exp(-iH_{\rm 2B}t/\hbar)$. 
We shall use  
$G(t,{\bf p},{\bf p}')=\langle {\bf p}|G(t){\cal S}|{\bf p}'\rangle$
in momentum space, where ${\cal S}$ projects onto the symmetrized relative 
momenta that correspond to two identical Bose atoms. At the initial time
$t_0$ the gas is assumed to be prepared in a (ideal gas) coherent state 
of the lowest energy mode 
\cite{initialstate}. 
The formal solutions then read:
\begin{align}
\nonumber
&\Phi({\bf p},t)=\int_{t_0}^{\infty} d\tau
\int d{\bf p}' G(t-\tau,{\bf p},{\bf p}')V({\bf p}')\Psi^2(\tau)\\
\nonumber
&+
\int_{t_0}^{\infty} d\tau
\int d{\bf q}'d{\bf p}'
G(t-\tau,{\bf p},{\bf p}_2')
[V({\bf p}_1')+V({\bf q}')]\\
&\times
\left[
\chi({\bf q}',{\bf p}',\tau)+2\sqrt{2\pi\hbar}^3
\delta({\bf q}')\Phi({\bf p}',\tau)\Psi(\tau)
\right]\Psi^*(\tau),
\label{Phi}
\end{align}
and
\begin{align}
\nonumber
&\Lambda({\bf q},{\bf p},t)=\int_{t_0}^{\infty} d\tau
\int d{\bf q}'d{\bf p}'
G(t-\tau,{\bf p},{\bf p}_2'-{\bf q}/2)\\
\nonumber
&\times\left[
\chi({\bf q}',{\bf p}',\tau)
+2\sqrt{2\pi\hbar}^3
\delta({\bf q}')\Phi({\bf p}',\tau)\Psi(\tau)
\right]\\
\nonumber
&\times[V({\bf p}_1'-{\bf q})+V({\bf q}'-{\bf q})]\Phi^*({\bf q},\tau)
e^{i\frac{q^2}{4m}(t-\tau)/\hbar}\\
\nonumber
&-\int_{t_0}^{\infty} d\tau\int d{\bf q}'d{\bf p}'
G(t-\tau,{\bf p},{\bf p}')\chi(-{\bf q},{\bf p}',\tau)
V({\bf q}+{\bf q}')\\
&\times
\left[
\Phi({\bf q}',\tau)+
\sqrt{2\pi\hbar}^3\delta({\bf q}')\Psi^2(\tau)
\right]^*e^{i\frac{q^2}{4m}(t-\tau)/\hbar},
\label{Lambda}
\end{align}
where ${\bf p}_1'={\bf q}'/2-{\bf p}'$ and 
${\bf p}_2'={\bf q}'/2+{\bf p}'$.
The form of Eqs.~(\ref{Phi}) and (\ref{Lambda}) is exact, 
save that terms that neither contribute to $g_2$ nor to $g_3$ are not 
given.
The arguments of the triple function $\chi$ for a uniform gas are
chosen here as the three body Jacobi momenta in 
Fig.~\ref{fig:Jacobi}.
 
The dynamic equation for $\chi$ involves the 
Hamiltonian of three atoms interacting pairwise,
\begin{align}
H_{\rm 3B}=-\frac{\hbar^2}{2m}(\Delta_{{\bf x}_1}+\Delta_{{\bf x}_2}
+\Delta_{{\bf x}_3})+V_{\rm 3B}, 
\end{align}
where
$V_{\rm 3B}=V({\bf r}_{23})+V({\bf r}_{13})+V({\bf r}_{12})$
(see Fig.~\ref{fig:Jacobi}). 
The corresponding symmetrized retarded Green's function in the three body 
center of mass frame is denoted here as 
$G_{\rm 3B}(t,{\bf q},{\bf p},{\bf q}',{\bf p}')$
in the representation through the Jacobi momenta in Fig.~\ref{fig:Jacobi}.
The only relevant contribution of the triple function reads 
\begin{align}
\nonumber
\chi({\bf q},{\bf p},t)=& 6\int_{t_0}^{\infty} d\tau
\int d{\bf q}'d{\bf p}'
G_{\rm 3B}(t-\tau,{\bf q},{\bf p},{\bf q}',{\bf p}_2')\\
&\times
V({\bf q}')\Phi({\bf p}',\tau)\Psi(\tau),
\label{chi}
\end{align}
where ${\bf p}_2'={\bf q}'/2+{\bf p}'$.

All nonlinear contributions in $\Psi$ and $\Psi^*$ up to
the fifth order are obtained by inserting Eqs.~(\ref{Phi}) and 
(\ref{Lambda}) into the right hand side of Eq.~(\ref{NLSEexact}) and then
replacing $\chi$ by Eq.~(\ref{chi}) and $\Phi$ by its leading first
contribution on the right hand side of Eq.~(\ref{Phi}). 
Equation (\ref{NLSEexact}) then becomes a non-Markovian 
NLSE \cite{TKKB} with the leading (two body) contribution
\begin{align}
\nonumber
i\hbar\frac{\partial}{\partial t}\Psi(t)=&(2\pi\hbar)^3
\int_{t_0}^\infty d\tau \
T(t-\tau,0,0)\Psi^2(\tau)\Psi^*(t)\\
&+\ldots,
\label{NMNLSE}
\end{align}
where $T(t)=V\delta(t)+VG(t)V$ is the Fourier transform of the energy 
dependent two body $T$ matrix 
\cite{Newton}. This time dependent transition matrix is sharply peaked
at $t=0$ with a width determined by the two body collisional duration.  
The fifth order nonlinear terms in $\Psi$ and $\Psi^*$
(three body contributions), 
indicated by the dots on the right hand side of Eq.~(\ref{NMNLSE}), 
involve up to four time integrals which have a similar form to the 
leading term.

To determine the equilibrium properties of the gas we shall assume first
that the scattering length is positive and the potential does not support
any bound clusters of atoms. In the zero temperature equilibrium 
the ansatz $\Psi(t)=\Psi\exp(-i\mu t/\hbar)$ then 
transforms the NLS Eq.~(\ref{NMNLSE}), including the three body
contributions, into its stationary form in the limit $t-t_0\to\infty$
\cite{timelimit}:
\begin{align}
\mu=g_2(\mu)|\Psi|^2+g_3(\mu)|\Psi|^4. 
\label{timeindependentNLSE}
\end{align}

The two body contribution to Eq.~(\ref{timeindependentNLSE}) is 
directly obtained
from Eq.~(\ref{NMNLSE}) through the Fourier transform of $T(t-\tau,0,0)$ 
at the energy $2\mu$ which yields the two body $T$ matrix
$T(2\mu,0,0)$ \cite{Tmatrix}. The low energy expansion 
\cite{squareroot,Adhikari}
\begin{align}
T(E,0,0)=\frac{1}{(2\pi\hbar)^3}\frac{4\pi\hbar^2}{m}a
\left[
1-i\sqrt{mE/\hbar^2}a+\ldots
\right]
\label{lowenergyT}
\end{align}
shows that the leading contribution to $\mu$ is the 
mean field chemical potential 
$\mu_0=4\pi\hbar^2 a |\Psi|^2/m$.
The three body contributions to the right hand 
side of Eq.~(\ref{NMNLSE}) 
(not shown explicitly)
can also be arranged in such a way that all bare
two body potentials are connected to two and three body $T$ matrices. 
The general form of the real part of Eq.~(\ref{timeindependentNLSE}) is then 
obtained as ($\eta=\sqrt{|\Psi|^2a^3}$):
\begin{align}
\nonumber
{\rm Re}\mu=&(2\pi\hbar)^3 {\rm Re} \ T(2\mu,0,0)|\Psi|^2\\
\nonumber
&+8\mu_0\left(
-3\sqrt{3}\ln\left|\frac{\sqrt{m2\mu}}{p^{(3)}}\right|
+4\pi\ln\left|\frac{\sqrt{m2\mu}}{p^{(4)}}\right|
\right)\eta^2\\
&+\frac{1}{2}(2\pi\hbar)^6T_{\rm 3B}^{(5)}(0,0,0)|\Psi|^4,
\label{remu}
\end{align}
where corrections smaller than $\mu_0\eta^2$ have been neglected.
The first term on the right hand side of Eq.~(\ref{remu}) is the 
two body part. All three body contributions 
to Eq.~(\ref{timeindependentNLSE}) are at least of third order in the 
two body $T$ matrix. The real parts of the third and fourth order terms 
exhibit a logarithmic divergence at low energies $\mu\to 0$ 
as indicated in the second term on the right hand side of Eq.~(\ref{remu}). 
The respective explicit expressions read
\begin{align}
\nonumber
&\ln\left|\frac{\sqrt{m2\mu}}{p^{(3)}}\right|
=-\frac{2\pi^5\hbar^4 m}{3\sqrt{3}a^4}
{\rm Re}\int \frac{d{\bf p} \ T(2\mu,{\bf p},0)}{2\mu-p^2/m+i0}\\
\nonumber
&\times
\sum_{\alpha =\pm1}
\bigg\{[T(2\mu,{\bf p},0)]^*\frac{
T(3\mu-\frac{3p^2}{4m},\alpha\frac{\bf p}{2},\frac{\bf p}{2})}
{2\mu-p^2/m-i0}\\
\nonumber
&\quad -[T(2\mu,{\bf p},0)]^*\frac{
T(\mu+\frac{p^2}{4m},\alpha\frac{\bf p}{2},\frac{\bf p}{2})}
{2\mu-p^2/m-i0}\\
&\quad +2T(2\mu,0,{\bf p})
\frac{
T(3\mu-\frac{3p^2}{4m},\alpha\frac{\bf p}{2},\frac{\bf p}{2})}
{2\mu-p^2/m+i0}\bigg\}
\label{thirdorder}
\end{align}
and
\begin{align}
\nonumber
&\ln\left|\frac{\sqrt{m2\mu}}{p^{(4)}}\right|
=\frac{\pi^4\hbar^4 m}{2a^4}
{\rm Re}\int 
\frac{d{\bf q}d{\bf p} \ T(0,0,{\bf q})T(0,{\bf p},0)}
{3\mu-\frac{q^2+p^2+{\bf q}\cdot {\bf p}}{m}+i0}\\
\nonumber
&\times
\sum_{\alpha,\beta =\pm1}
\bigg\{\frac{
T(-\frac{3q^2}{4m},\alpha\frac{\bf q}{2},{\bf p}+\frac{\bf q}{2})
T(-\frac{3p^2}{4m},{\bf q}+\frac{\bf p}{2},\beta\frac{\bf p}{2})}
{(2\mu-q^2/m-i0)(2\mu-p^2/m+i0)}\\
&+2\frac{
T(-\frac{3q^2}{4m},\alpha\frac{\bf q}{2},{\bf p}+\frac{\bf q}{2})
T(-\frac{3p^2}{4m},{\bf q}+\frac{\bf p}{2},\beta\frac{\bf p}{2})}
{(2\mu-q^2/m+i0)(2\mu-p^2/m+i0)}\bigg\}.
\label{fourthorder}
\end{align}
The logarithmic behavior of the real parts of
the integrals in Eqs.~(\ref{thirdorder}) 
and (\ref{fourthorder}) was determined with the 
methods in Ref.~\cite{Adhikari}. The momentum units $p^{(3)}$ and $p^{(4)}$ 
are found using the regular parts of the integrals in the limit $\mu\to 0$.

The last contribution to the right hand side of Eq.~(\ref{remu}) involves
the $T$ matrix for three asymptotically free incoming and outgoing 
atoms, $T_{\rm 3B}(E)=V_{\rm 3B}+V_{\rm 3B}G_{\rm 3B}(E)V_{\rm 3B}$. 
The regular part of $T_{\rm 3B}$, denoted here by $T_{\rm 3B}^{(5)}$, 
consists of all multiple scattering contributions to $T_{\rm 3B}$ from the 
fifth order upward \cite{Faddeev}. $T_{\rm 3B}^{(5)}(E)$ can be evaluated 
at $E=0$ 
\cite{Adhikari}
and is the only contribution to $g_3$ that depends on the complete    
zero energy three body scattering state.

Equation (\ref{timeindependentNLSE}) also exhibits an imaginary part 
\begin{align}
{\rm Im}\mu=\mu_0
\left[
-\sqrt{\frac{m2\mu}{\hbar^2}}a+\frac{8\pi\hbar}{\sqrt{m2\mu}a}\eta^2
+{\cal O}(\eta^2\ln\eta)
\right]
\label{immu}
\end{align}
whose first leading term stems from Eq.~(\ref{lowenergyT}). 
The second term is the 
leading contribution to the imaginary part of the integral in
Eq.~(\ref{thirdorder}). The imaginary part of $\mu$ describes the
exchange of atoms between the condensed and thermal fraction 
\cite{TKKB}
and is thus required to vanish exactly in equilibrium as soon as all few body 
scattering contributions are taken into account. The right hand sides of 
Eqs.~(\ref{remu}) and (\ref{immu}) change by an amount smaller than 
$\mu_0\eta^2$ if $\mu$ is replaced by its leading contribution $\mu_0$.
The square root terms in Eq.~(\ref{immu}) then, indeed, cancel each other
and Eqs.~(\ref{remu}) and (\ref{timeindependentNLSE}) yield 
\begin{align}
\nonumber
g_2=&(2\pi\hbar)^3{\rm Re} \ T(2\mu_0,0,0),\\
\nonumber
g_3=&\frac{32\pi\hbar^2}{m} a^4
\left(
-3\sqrt{3}\ln\left|\frac{\sqrt{m2\mu_0}}{p^{(3)}}\right|
+4\pi\ln\left|\frac{\sqrt{m2\mu_0}}{p^{(4)}}\right|
\right)\\
&+\frac{1}{2}(2\pi\hbar)^6T_{\rm 3B}^{(5)}(0,0,0).
\label{result}
\end{align}

A trapped dilute BEC can usually be divided into regions with an extent 
much larger than the range of the inter-atomic forces but still much smaller
than the scale of the spatial variation of the condensate wave function
$\Psi({\bf x})$. In these regions the gas can be considered as locally 
homogeneous and the coupling constants in Eq.~(\ref{result}) remain valid
when $\mu_0$ is replaced by the local chemical potential 
$\mu_0({\bf x})=4\pi\hbar^2a|\Psi({\bf x})|^2/m$. 

A realistic molecular potential $V({\bf r})$ usually
supports two body bound states.
The coupling constant $g_3$ then acquires   
an imaginary part through $T_{\rm 3B}^{(5)}$. According to three body 
collision theory this imaginary part corresponds to the threshold-less 
formation of dimer molecules and was determined here via an optical theorem 
with the methods of 
Ref.~\cite{Sandhas}. We thus obtained $2 {\rm Im}g_3/\hbar=-K_3/6$ 
where $K_3$ is the 
three body recombination rate constant for an ultra-cold non condensed 
Bose gas. This was 
determined in Ref.~\cite{Moerdijk} using classical probability arguments, 
by assuming that three atoms are lost in each event. The factor 
of $1/6$ expresses the fact that all three condensed atoms share the same 
quantum state
\cite{Kagan} 
and was verified in subsequent experiments 
\cite{Burt}.
In the presence of bound states the condensate can only assume a metastable
state. The change of its density, however, is negligible on time scales
comparable to collisional durations so that the present results remain 
valid. 

If $a$ is negative but the trapped gas is still stable 
with respect to collapse the above analysis slightly modifies:
$\mu_0({\bf x})$ then becomes negative and instead of Eq.~(\ref{immu}) one 
obtains real contributions, $\mu_0\sqrt{m2|\mu|}a/\hbar$ and
$\mu_0 8\pi\hbar|\Psi|^2 a^2/\sqrt{m2|\mu|}$, through Eq.~(\ref{lowenergyT})
and the integral in Eq.~(\ref{thirdorder}), respectively. 
The contribution of Eq.~(\ref{thirdorder}), with $|\mu|$ replaced by 
$|\mu_0|$, revises the two body coupling constant to
$g_2=(2\pi\hbar)^3T(2\mu_0,0,0)-4\pi\hbar a^2\sqrt{2|\mu_0|/m}$.  
This leads to $g_2=\frac{4\pi\hbar^2}{m}a[1+{\cal O}(\eta^2)]$ 
by Eq.~(\ref{lowenergyT}). If the gas collapses 
\cite{Donley}
the equilibrium assumption as well as the subsequent concept of two and 
three body coupling constants break down. 

When a three body bound state emerges close to the zero energy 
threshold $g_3$ is dominated by the residue of the corresponding pole through 
$T_{\rm 3B}^{(5)}$
\cite{Adhikari}.
The real part of $g_3$ can then assume all positive and negative
values in complete analogy to the two body case. This phenomenon was 
discussed in Ref.~\cite{BraatenBulgac} in connection with the 
Efimov effect in the three body energy spectrum
\cite{Efimov}.

The three body coupling constant in Eq.~(\ref{result}) has the form 
predicted by Wu
\cite{Wu}
and Braaten et al.~\cite{BraatenNieto,BraatenBulgac}
and can be determined quantitatively 
by solving the two and three body scattering problem in vacuum
\cite{Goey}. 
The explicit representation in Eq.~(\ref{result})
also allows to directly relate all our current 
knowledge about two and three body transition amplitudes to properties of
the gas. The present
theory is valid for all (non singular) molecular potentials and
avoids divergences. Convergence at high energies has been achieved by taking
into account the exact binary potential which enters the coupling constants 
through two and three body $T$ matrices. At low energies the arguments 
of the infrared divergent contributions to $g_3$ have been determined to
multiples of $\mu_0$ which is the energy scale that corresponds to the healing
length
\cite{Dalfovo}.

In summary, we have derived the explicit three body contact potential
for a dilute condensed Bose gas from a microscopic many body theory.
The respective coupling constant, $g_3$, that reflects general equilibrium
properties of an interacting BEC has been expressed in terms of the 
binary potential through two and three body transition matrices. Its 
explicit form allows for a determination of the magnitude of $g_3$ 
from few body physics. We have derived the heuristic three body recombination 
rate constant of Ref.~\cite{Moerdijk} as well as phenomena related to 
resonant three body bound states.

I am greatly indebted to Thomas Gasenzer, Simon Gardiner, Robert Roth,
and Keith Burnett for valuable discussions. 
This research was supported by the Alexander von
Humboldt Foundation and the UK EPSRC.  


\begin{thebibliography}{99}
\bibitem{timelimit}
For the density $|\Psi|^2$ to become stationary
it is sufficient that $t-t_0$ by far exceeds the two and three body 
collisional durations. After this time the gas has
lost its memory of the particular initial state.
\bibitem{Tmatrix}
The energy argument is then
approached from the upper half of the complex plane, i.e.~$T(E)=T(E+i0)$.
\bibitem{squareroot}
Here and throughout this article we choose the branch of the complex square 
root with a positive imaginary part.
\end{thebibliography}
\end{document}